\documentclass[showpacs,preprintnumbers,amsmath,amssymb,preprint]{revtex4}
\usepackage[final]{graphicx}
\usepackage[english]{babel}
\begin{document}
\title{ Suppression of $\pi^0$ production at large $p_\perp$ in
central $Au +Au$ collisions at $\sqrt{s_{NN}} = 200$ GeV and
quark-gluon plasma}
\author{Yu.A. Tarasov}
\email{tarasov@dni.polyn.kiae.su} \affiliation{
 Russian ResearchF
Center ''Kurchatov Institute'', 123182, Moscow,  Russia }

\author{ S.L. Fokin}
\email{tarasov@dni.polyn.kiae.su} \affiliation{
 Russian ResearchF
Center ''Kurchatov Institute'', 123182, Moscow,  Russia }
\date{\today}

\begin{abstract}
We investigate the suppression of $\pi^0$-spectrum in a wide range
of $p_\perp$ up 60 GeV/c which is caused by the energy loss of the
gluon and quark jets in quark-gluon plasma. The physical
characteristics of initial and mixed phases were found in the
effective quasiparticle model by analogy with previous
work~\cite{5}. The PHENIX data up 10 GeV/c within the limits of
precision are described by quasiparticle model with decrease of
the thermal gluon mass and effective coupling in the region of
phase transitin (at $T\to T_c$ from above). We also take into
account the intrinsic transverse momentum $k_\perp$ of partons.

The suppression factor $R_{AA}(p_\perp)$ shows the weak rise with
increase of $p_\perp$ above 4 GeV/c, then it reaches smooth
maximum at $p_\perp \sim 20$ GeV/c and then decrease  at
$p_\perp\sim 60$ GeV/c again to value $R_{AA}$ at $p_\perp\simeq$
4 GeV/c. The factor $R_{AA}$ in this range of $p_\perp$ is
changihg weakly if intrinsic momentum $k_\perp$ is taken into
account.

\end{abstract}


\maketitle

\section{Introduction    \label{sec1}}

Energy loss of high energy gluon and quark jets in relativistic
$A+A$ collisions leads to jet quenching and to suppression of
hadron spectra and thus probes the quark-gluon plasma. Recently
the considerable suppression of $\pi^0$ spectra was found in
central $Au+Au$ collisions at $\sqrt{s}$=200 GeV~\cite{1}.

The energy loss was investigated, for example, in Ref.~\cite{2} in
various orders in opacity $L/\lambda$ (where L is nuclear radius
and $\lambda$ is the gluon mean free path). It was shown that
opacity expansion is strongly dominated by the first term. It was
shown also in Ref~\cite{3} that the finite kinematic boundaries
decrease the energy loss as compared to the asymptotic limits.
Recently  the suppression of $\pi^0$ spectra in central $Au+Au$
collisions at $\sqrt{s}$=130 GeV was investigated in Ref~\cite{4}
with accounting of dominant first term and of the finite kinematic
boundaries. In this work
 the physical characteristics
of initial and mixed plasma phase (i.e. the values
$T_0,\tau_0,\tau_c$) were used which have been found in
Ref~\cite{5} on the basis of the quasiparticle model and
isentropic expansion. We find that the suppression one can
describe by quasiparticle model with decrease of the thermal gluon
mass and effective coupling in a region of phase transition (at $T
\to T_c$ from above). The possibility of hot glue production at
the first stage was also considered in Ref~\cite{4}, The gluon
density was calculated with accounting of plasma expansion and it
was shown that plasma is sufficiently thin : the average number of
jet scatterings $\bar n\simeq$ 1.3, so it is possible to use the
model of the single hard medium-induced scattering. The physical
characteristics of plasma at $\sqrt{s}$=200 GeV was found by
analogy with the Ref.~\cite{5}. We have used the values $T_c$=177
MeV and $\mu_B$=29 MeV for 200 GeV $Au+Au$ central
collisions~\cite{6}. These values are consistent with data for
$\bar p/p$ ratio and for the number of net protons ($\simeq$ 5) at
midrapidity~\cite{7}. Using the formulas (10-18) in
Ref.~\cite{4},we have found for gluons
\mbox{$\left(\frac{dN^{g}}{dy}\right)_{y=0}\simeq 1070$} and for
hot glue the temperature $T_g \simeq$ 373 MeV. We have obtained
also the initial gluon density \mbox{$n_g(T_g)\simeq 24.77\;
m_{\pi}^3$}. The time which is required to achieve the equilibrium
for gluons is:
\begin{equation}
\label{eq.1} \tau_g = \frac{dN_g}{dy\pi R_{Au}^{2}n_g} \simeq
\frac{0.587}{m_{\pi}}=0.834 \;\; \textrm{fm.}
\end{equation}
 The equilibrium temperature $T_0$ for both quarks and gluons and
corresponding time $\tau_0$ can be found using the formulas
(14-23) in Ref.~\cite{5}. We have at $\sqrt{s}=200$ GeV: $T_0$
=224 MeV, $\tau_0 \simeq 1.585/m_{\pi} \simeq 2.26$ fm. The
corresponding entropy density is $s_0(T_0) = 58.98\; m_{\pi}^3$
and total entropy is $S_0 = S_0(T_0)V_0 \simeq 6850$. We have also
at $T=T_c=177$ MeV: $s_c(T_c)\simeq 16.72\; m_{\pi}^3$ and $\tau_0
\simeq 4.5/m_{\pi}$
 (i.e. $\simeq  6.4$ fm).
 In Ref.~\cite{5} it was shown, that from conservation of entropy
 and  number of net nucleons follows, that the massive constituent
 quarks ($m_q$ and $m_s$) appears with decrease of number of
 degrees of freedom in the presence of octet of pseudogoldstone
 states. In addition with the same effective number of degrees of
 freedom appears hadrons and resonances in hadron part of mixed
 phase (for SPS and RHIC at $\sqrt{s}=130$ GeV). One can show that
 it is completely fulfilled for RHIC at $\sqrt{s} = 200$ GeV.

 In Sec.~\ref{sec2} we calculate the energy loss $\Delta E$ of the
 high energy gluon and quark jets with energy $E$ in expanding
 quark-gluon plasma (at $Au+Au$ collisions at $\sqrt{s}=200$ GeV)
 in the dominant first order of opacity expansion(by analogy with the
 Ref.~\cite{4}) We use here the characteristics of expanding
 plasma which are found in effective quasiparticle model (with
 accounting of the hot glue production at first stage).

 In Sec.~\ref{sec3} we calculate the suppression of $\pi^0$
 spectra  in central $Au+Au$ collisions at $\sqrt{s}=200$ GeV in
 wide range of $p_{\perp}$ ($4\le p_{\perp}\le 60$ GeV). In this
 Sec. we do not take into account the intrinsic transverse
 momentum $k_{\perp,i}$ of partons.

 In Sec.~\ref{sec4} we calculate  the suppression of $\pi_0$
 spectra taking into account the intrinsic transverse momentum of
 partons. We use here a Monte-Carlo method for calculation
 of invariant cross section of hadron production. We show, that if
 take into consider the physical restrictions,then the parton cross
 section does not have the divergence at too large
 $k_{\perp,i}$, which are considered in other works (for example,
 in Ref.~\cite{8}). Therefore there is no necessity to introduce
 a regulator $\mu^2$ in the denominators of the parton-parton
 cross sections. We show that this model reproduces the data for
 $\pi_0$ spectra at high $p_{\perp}$ in pp-collisions at
 $\sqrt{s}=200$ GeV without introduction of $K$-factor ( at
 $<k_{\perp}^2> \simeq 1.8$ GeV$^2$).   We show also, that
 suppression factor $R_{AA}(p_{\perp})$ is changes weakly in
 above-mentioned range of $p_{\perp}$, if intrinsic momentum
 $k_{\perp,i}$ is taken into account.

 In Sec.\ref{sec5} - Conclusion.

\section{Energy loss of high energy gluon and quark jets  \label{sec2}}

In previous section  the physical characteristics of the plasma
stage for RHIC at $\sqrt{s}=200$ GeV were found. The production of
hot glue at the first stage is caused by relatively large $gg$
cross section in comparison with $qg$ and $qq$ cross section. We
use these characteristics for investigation of energy loss of the
high energy parton jets. The dominant first order radiation
intensity distribution $\frac{dI}{dx}$ for expanding plasma is
given in Ref.~\cite{3} and correspond to formula~(\ref{eq.3}) in
Ref.~\cite{4}, which have the  form:
\begin{eqnarray}
\label{eq.2} &&\frac{dI}{dx} = \frac{9 C_{R} E}{\pi^2}
\int\limits_{z_0}^{\infty}dz\,\rho (z)
\int\limits_{|k|_{\min}}^{|k|_{\max}} d^{2} {\textbf k}\ ,
\alpha_{s} \nonumber \\ & &  \int\limits_0^{q_{\max}} d^{2}
{\textbf q} \frac{\alpha_{s}^2}{[{\textbf q}^{2}+{\mu}^{2}(z)]^2}
\frac{{\textbf k} {\textbf q}}{{\textbf k}^{2}( {\textbf k} -
{\textbf q})^2}\ \left[1 - \cos\frac{( {\textbf k} - {\textbf
q})^{2}(z-z_0)}{2x(1-x)E}\right] \,,
\end{eqnarray}
 In this formula  the value $E$ is the jet energy and $C_R$ is the color
 factor of jet:
 $C_R$ is $N_c$ for gluon and $\frac{4}{9} N_c$ for quark jet. It is
assumed that quark-gluon plasma can be modeled by the well
separated color-screened Yukawa potentials. In Ref.~\cite{4} was
shown, that this condition is well realized for partons scattering
on gluon potential, but this is not realized for partons
scattering on the quark Yukawa potential. Therefore the estimation
of parton energy loss $\Delta E$ on the gluon Yukawa potential is
the most real in this model. In formula~(\ref{eq.2}) we have
$|q|_{\max}= \sqrt{3\mu(\tau)E}$, where
$\mu^2(\tau)=4\pi\alpha_{s}T^{2}(\tau)$. and the kinematic bounds
are ${\textbf k}_{\max}^2 = \min{\{4E^{2}x^2,4Ê^{2}x(1-x)\}}$ and
${\textbf k}_{\min}^2 = {\mu}^{2}(\tau)$ for gluons with the light
cone momentum fraction x. The value $z=\tau$ is limited in fact by
duration of the plasma phase $\tau_g \le\tau \le\tau_c$ (and
$\tau_c \le R_{Au}$). The value $\rho(\tau)$ is gluon density
$n_g$ at time $\tau$ along the jet path, which is calculated in
effective quasiparticle model. The energy loss $\Delta E_g$ of
gluon jet is determined by integration of $dI/dx$ in~(\ref{eq.2})
over x. The corresponding integral $I_{0}(E,\tau)$ (look a
formula~(\ref{eq.22}) in Ref.~\cite{4}) is calculated in
above-mentioned finite kinematic bounds by Monte Carlo method for
values $\tau(T)$ from $\tau_g$ to $\tau_c$ and for of $E$ up to
$60-70$ GeV. The energy loss $\Delta E_g$ is determined by
formula:
\begin{equation}
\label{eq.3}
 \Delta E_g(E) = \frac{9C_{R}E}{\pi^2}
\int\limits_{\tau_g}^{\tau_c} \frac{d\tau}{{\mu}^{2}(\tau)}
I_{0}(E,\tau) n_{g}(\tau).
\end{equation}
\begin{figure} [ht]
\centering \mbox{\includegraphics*[scale=1]{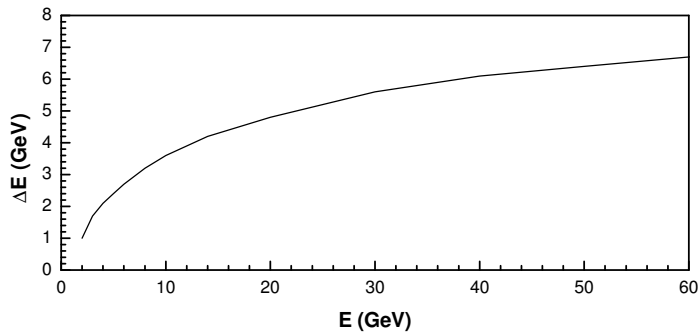}}
\caption{The energy loss of gluon jet in quark-gluon plasma at RHIC
energy ($\sqrt{s}=200$ GeV) \label{Fig.1}}
\end{figure}

\begin{figure} [ht]
\centering \mbox{\includegraphics*[scale=1]{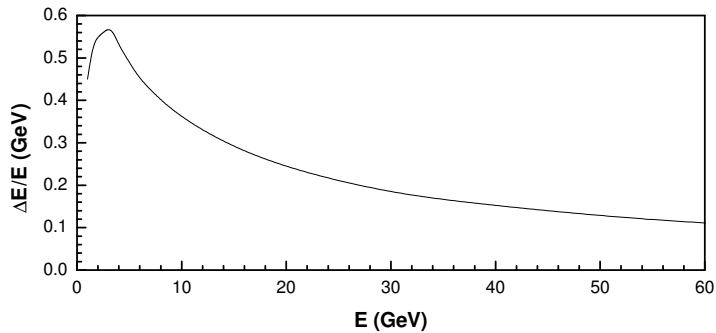}}
\caption{The relative energy loss of gluon jet in quark-gluon plasma. \label{Fig.2}}
\end{figure}
The values $I_{0}(\tau)$ and $n_{g}(\tau)$ for expanding plasma at
$\sqrt{s} = 200$ GeV and the values of energy loss for gluon jet
we show in Table~\ref{Tab.1} for example at $E = 10$ GeV. The
complete energy loss $\Delta E$ is calculated by numeral
integration over $\tau$. We have $\Delta E = 3.6$ GeV at $E = 10$
GeV, i.e. not great increase in comparison with $\sqrt{s}=130$
GeV. The complete energy loss  $\Delta E_{g}(E)$  of gluon jet is
shown in Figure~\ref{Fig.1} and the relative energy loss $\Delta
E_g/E$ in Figure~\ref{Fig.2}. For energy loss of quark jet we have
$\Delta E_{q}{E}= \frac{4}{9}\; \Delta E_g(E)$. That is important
non-Abelian feature of parton energy loss~\cite{9}.

\begin{table}
\caption{The physical values for expanding plasma at $\sqrt{s} =
200$ GeV and energy loss $\Delta E_g$ at $E = 10$ GeV~\label{Tab.1}}.
\begin{tabular}{ccccccc}
\hline \hline
$T$ (MeV) & $\tau$ ($m_{\pi}^{-1}$)  & $\alpha_s$ &
$\mu^{2}$ $(m_{\pi})$ & $n_g$ ($m_{\pi}^{3}$) & $I_0$ & $\Delta E$
(GeV)
\\ \hline
373 & 0.587 & 0.269 & 24.34 & 24.77 & 0.097  & 2.64
\\
350 & 0.64 & 0.268 & 21.35 & 20.12 & 0.095 &  2.61
\\
325 & 0.713 & 0.266 & 18.27 & 15.75 & 0.092 & 2.36
\\
300 &0.808 &  0.263 &  15.39 & 12.035 & 0.089 &  1.905
\\
275 & 0.935 & 0.259 &  12.74  & 8.92  &  0.084  & 1.62
\\
250 & 1.116 & 0.254 &  10.32  & 6.34  &  0.081  & 1.34
\\
224 & 1.585 & 0.246 &  8.024  & 3.87  &  0.08  &  1.065
\\
210 &  1.935 & 0.239 & 6.85  &  2.99  &  0.078  &  0.96
\\
200  &  2.293 & 0.231 & 6.01 &  2.428  &  0.076 &  0.84
\\
190  &  2.838 & 0.22  & 5.16  & 1.886  &  0.07  &  0.7
\\
185  &  3.267 & 0.21  & 4.67  & 1.612  &  0.063 &  0.59
\\
180  &  3.98  & 0.19  & 4.025  & 1.32  & 0.05  &  0.448
\\
177  &  4.5  & 0.057  &  1.16  &  1.39 & 0.00063 & 0.0204
\\   \hline \hline
\end{tabular}
\end{table}
It should be noted, that these energy loss correspond to the model
of phase transition with decrease of effective coupling strength
$G(T)$ and gluon mass $m_{g}(T)$ for $T \to T_c$ from above. In
Ref.~\cite{5} we already noted, that model with increase of $G(T)$
for $T \to T_c$ agrees with new lattice data and provides
description of baryon and meson spectra, but it give too large
suppression of hadrons with large $p_{\perp}$ which disagrees with
data.

\section{  Suppression of pions with large \mbox{$p_{\perp}$}
at \mbox{$\sqrt{s}=200$} GeV/c without accounting  of intrinsic
momenta \mbox{$k_{\perp}$} of partons. \label{sec3}}

The energy loss reduces the jet energy before fragmentation, where
the jet transverse momentum is shifted on the value $\Delta E(E)$.
If this effect is taken into account, we should replace the vacuum
fragmentation function by the effective one
$\frac{z_c^{\ast}}{z_c}D_{h/c}(z_c^{\ast},{Q^{\ast}}^{2})$~\cite{10}
where
\begin{equation}
\label{eq.4} z_c^{*} = \frac{z_c}{1 -\frac{\Delta E (E)}{E}}.
\end{equation}
The invariant cross section of hadron production in central $A+A$
collisions is given by
\begin{eqnarray}
\label{eq.5} &&E_{h}\frac{d\sigma_{h}^{AA}}{d^3p} =
\int\limits_0^{b\max}d^2b\; d^2r \,t_{A}(r) t_{A}(|\textbf  b -
\textbf r|) \sum_{abcd}\int
dx_{a}dx_{b}d^{2}k_{\perp,a}d^{2}k_{\perp,b} \times \nonumber
\\&\times& g_{A}(k_{\perp,a},Q^{2},r) g_{A}(k_{\perp,b},Q^2,|\textbf
b -\textbf  r|) f_{a/A}(x_{a},Q^2,r)f_{b/A}(x_b,Q^2,|\textbf  b -
\textbf r|)\times \nonumber
\\&\times&\frac{d\sigma}{d\hat t} \frac{z_{c}^{*}}{z_c}
\frac{D_{h/c}(z_{c}^{*},\hat Q^2)}{\pi z_c}\,.
\end{eqnarray}
Here $t_{A}$ is the nuclear thickness function, the $k_{\perp,a}$
and $k_{\perp,b}$ are the intrinsic transverse momenta of partons,
the values $f_{a/A}$ and $f_{b/A}$ are the partons structure
function. It is usually assumed that distribution
$g_{A}(k_{\perp})$ has a Gaussian form. It should be noted, that
intrinsic $k_{\perp}$ are more important for final hadron spectra
for SPS energies. With the increase of energy the spectra become
flatter and intrinsic momenta are less important~\cite{8}. At
higher energy the intrinsic momenta $k_{\perp}$ are included in
order to account for phenomenologically the next -to- leading
order correction~\cite{11} (instead of K-factor), The accounting
of $k_{\perp}$ change weakly the factor $R_{AA}$. We will take
into account the intrinsic momenta of partons in the next Section.

The upper limit for the impact parameter is $b_{\max}\simeq
0.632\; R_{Au}$ for $10\%$ central $Au+Au$ collisions. Since the
energy of
 collisions is very high, we will consider for simplicity the
 collisions of flat disks. In that case the integral of
 overlapping is $T_{AA}(b) = \frac{A^2}{\pi R_A^2} \simeq 262$
 fm$^{-2}$.  The parton shadowing
 factor $S_{A}$ for flat disks also should be taken into account.We used the shadowing functions
 $S_{A}(x,Q^2$) from EKS98 parameterization~\cite{12}. These
 shadowing functions affected weakly on suppression factor
 $R_{AA}$, especially at large $p_{\perp}$ (at least at
 $p_{\perp}< 20$ GeV/c).It was shown also in Ref.~\cite{17}.
  At more of $p_{\perp}$ it is possible the
 additional decrease of $R_{AA}$ due to EMC modification of
 nuclear structure function.
 We use the parton distribution from Ref.~\cite{13} and fragmentation
function $D_g$ and $D_q$ from Ref.~\cite{14} both in $LO$
parametrization.

We will use the equation (\ref{eq.5}) in another aspect, in order
to compare the energy loss with other works, for example with
Ref.~\cite{15} and~\cite{16}. Let us consider at first the hadrons
production at parton collisions in the absence of the medium. A
parton produced initially with transverse momentum $p_{\perp} + u$
fragmentire into hadron with momentum $p_{\perp}$. We have $z =
\frac{p_{\perp}}{p_{\perp}+u}$, and we use the factorization scale
$Q = \frac{p_{\perp}}{2z}$, that is $Q(u) =\frac{p_{\perp}+u}{2}$.
We have here $u_{\min} = 0$ and $u_{\max} = \frac{\sqrt{s}}{2} -
p_{\perp}$. The invariant cross section for $\pi^0$ production in
$pp$ collisions for example gluon jet is (without accounting of
partons intrinsic momentum):
\begin{eqnarray}
\label{eq.6} E_{\pi}\frac{d\sigma^{pp}}{d^{3}p} = \frac{9}{2}
\int\limits_0^{\frac{\sqrt{s}}{2}-p_{\perp}} du
\frac{\alpha_s^{2}(u)}{p_{\perp}(p_{\perp}+u)^4} D(z(u),Q^{2}(u)) \times \nonumber \\
\int\limits_{\frac{1}{1-k(u)}}^{\frac{1}{k(u)}} d\xi f_{gg}(\xi)
x_{1}G_{g}(x_{1},Q) x_{2}G_{g}(x_{2},Q),
\end{eqnarray}
 where $k(u)=\frac{p_{\perp}+u}{\sqrt{s}}$. We introduce here the
new variable $x_1 = \frac{(p_{\perp}+u)\xi}{\sqrt{s}}$, $x_2 =
\frac{(p_{\perp}+u)\xi}{\sqrt{s} (\xi-1)}$. We describe the
contribution of the different elementary sections in variable
$\xi$:
\begin{equation}
\label{eq.7} f_{gg}(\xi) = \frac{3(\xi-1)}{\xi^4} -
\frac{(\xi-1)^2}{\xi^6} + \frac{1}{(\xi-1)\xi^3} +
\frac{(\xi-1)^2}{\xi^3},
\end{equation}
\begin{equation}
\label{eq.8} f_{qg}(\xi) = \frac{4}{9} \left(\frac{1}{\xi^2} +
\frac{1}{\xi^3} - \frac{1}{\xi^4}\right) +
\frac{\xi^2+1}{\xi^4(\xi-1)} + \frac{\xi-1}{\xi^2}
+\frac{(\xi-1)^3}{\xi^4}.
\end{equation}
 For corresponding contributions into quark-quark cross section for
identical $f_{qq}^{id}$ and for different $f_{qq}^{dif}$ quarks we
have:
\begin{equation}
\label{eq.9} f_{qq}^{id} = \frac{4}{9}
(\frac{\xi^2+1}{\xi^4(\xi-1)} + \frac{\xi-1}{\xi^2}
\frac{(\xi-1)^3}{\xi^4}) - \frac{8}{27} \frac{1}{\xi^2},
\end{equation}
\begin{equation}
\label{eq.10} f_{qq}^{dif} = f_{qq}^{id}+ \frac{8}{27}
\frac{1}{\xi^2}.
\end{equation}
In the presence of medium the parton losses the additional energy
$\Delta E$ and we have from~(\ref{eq.4}):
\begin{equation}
\label{eq.11} z^{*}(u,\Delta E) =
\frac{p_{\perp}}{p_{\perp}+u-\Delta E}
\end{equation}
and also:
\begin{equation}
\label{eq.12} Q^{*}(u) = \frac{p_{\perp}}{2z^{*}} =
\frac{p_{\perp}+u-\Delta E}{2}.
\end{equation}
The value $z_{\max}^{g}$ for gluons (i.e. and $u_{\min}^g$) in the
presence of medium we  find from relation (\ref{eq.4}:
${z^{*}_{\max}}^g= \frac{p_{\perp}}{E- {\Delta E}^g}= 1$ for every
value of $p_{\perp}$ (Ref.~\cite{4}). The value $z_{\max}^{q}$ for
quarks is fined by substitution of $(\Delta E)^{q}$ instead of
$(\Delta E)^{g}$.
 The value $E_{\min}$ which correspond to gluon jet for
given $p_{\perp}$ can be found from Figure~\ref{Fig.1} (in large
scale). In the range $E_{\min}\le E \le \frac{\sqrt{s}}{2}$ we
construct the dependence $(\frac{\Delta E}{E})^{g}(z)$ using
Figure~\ref{Fig.2}. For analogous function for quarks
$(\frac{\Delta E}{E})^{q}(z)$ we use the relation $(\Delta
E)_{q}=\frac{4}{9} (\Delta E)_{g}$. From these relations we find
the values $(\Delta E)_{g,q}(u)$. For example, we find for gluons
at $p_{\perp}=8$ GeV (with $(3-5)\%$ precise):
\begin{equation}
\label{eq.13} \left(\frac{\Delta E}{E}\right)^{g}(z)=0.42z + 0.05
+0.35(z-0.08)(0.665-z).
\end{equation}
The corresponding equation for quarks is (with accounting that
$z_{\max}^{q} \ne z_{\max}^{g}$):
\begin{equation}
\label{eq.14} \left(\frac{\Delta E}{E}\right)^{q} = \frac{4}{9}
\left(0.4z +0.049 + 0.3(z-0.08)(0.835-z)\right).
\end{equation}
At increase of $p_{\perp}$ the dependence $(\Delta E)/E$ become
linear. For example, at $p_{\perp} = 20$ GeV we have:
\begin{eqnarray}
\label{eq.15} \left(\frac{\Delta E}{E}\right)^{g}(z) &=& 0.227z +
0.036\; (\textrm{with}\; 0.2\le z\le  0.786), \nonumber \\
\left(\frac{\Delta E}{E}\right)^{q}(z)& =& \frac{4}{9}
\left(\frac{\Delta E}{E}\right)^{g}  (\textrm{with}\; 0.2\le z\le
0.893).
\end{eqnarray}

We have here in variable u:
\begin{equation}
\label{eq.16} (\Delta E)^{g}(u) = 5.26 + 0.036 u \;
(\textrm{with}\; 5.4\le u \le 80\; (\textrm{GeV})).
\end{equation}

 The fragmentation scale is now:
\begin{eqnarray}
\label{eq.17} Q_{g}(u) &=& \frac{p_{\perp}+u-\Delta E^{g}(u)}{2}, \nonumber\\
Q_{q}(u) &=& \frac{p_{\perp}+u-\Delta E^{q}(u)}{2}.
\end{eqnarray}

Let us write down  the formula of type~(\ref{eq.6}) in $AA$
collisions for $\pi^0$ production in gluon jet with accounting of
energy loss in medium:
\begin{eqnarray}
\label{eq.18} E_{\pi}\frac{d\sigma^{AA}}{d^{3}p} &=& \frac{9}{2}
\; \;\frac{A^2}{\pi
R_{A}^2}\int\limits_{u_{\min}(p_{\perp})}^{\frac{\sqrt{s}}{2} -
p_{\perp}} du
\frac{\alpha_{s}^{2}(Q(u))}{p_{\perp}(p_{\perp}+u)^4}
\frac{D_{g}(z^{*}(u),Q_{g}(u))}{1- \frac{\Delta
E^{g}(u)}{p_{\perp}+u}}\nonumber
\\ && \int\limits_{\frac{1}{1-k(u)}}^{\frac{1}{k(u)}} d\xi
f_{gg}(\xi) x_{1}G_{g}(x_1,Q(u)) x_{2} G_{g}(x_2,Q(u)).
\end{eqnarray}
For calculation of complete energy loss we must take into account
also quark-gluon and quark-quark collisions with corresponding
structure function and elementary cross section
(\ref{eq.7}-\ref{eq.10}) and the energy losses $(\Delta
E)^{g,q}(u)$. We take into account also the gluon collisions with
sea quarks.

The suppression factor $R_{AA}(p_{\perp})$ is the ratio of the sum
of invariant cross section of type (\ref{eq.18}) to the sum cross
section of binary  NN collisions of type(\ref{eq.6}) without
accounting of nuclear effects - jet quenching and shadowing. We
have calculated the factor $R_{AA}(p_{\perp})$ in wide range of
$p_{\perp}$ from 4 up 60 GeV/c. In order to avoid too small values
at calculation for great $p_{\perp}$, we multiply the numerator
and denominator in this ratio on value $p_{\perp}^4$.

It should be noted, that energy loss in medium can not be
expressed by simple manner through vacuum cross section
$\frac{d{\sigma}(p_{\perp}+\Delta E)}{dp_{\perp}^2}$ (as for
example in~\cite{15}, \cite{16}). The situation have more complex
character.

We show in Table~\ref{Tab.2} the results for suppression factor
$R_{AA}(p_{\perp})$ for flat disks with accounting of nuclear
shadowing function $S_{A}(x,Q^2)$ from EKS98
parameterization~\cite{12}. We take into account in this Table
also the intrinsic transverse momentum of partons (look at
formula~(\ref{eq.24}) in~\ref{sec4}). The contribution of nuclear
shadowing effects in $R_{AA}(p_{\perp})$ is small (less than
10\%). It was noted also in the work~\cite{17}. At large
$p_{\perp}>20$ GeV this contribution increase up to $\sim 20\%$
and that gives decrease of $R_{AA}$ due to EMC effect.

It should be noted, that at large $p_{\perp}$ the value
$R_{AA}(p_{\perp})$ is determined quantitatively in the main by
quark jets --- the contribution of gluons into numerator of ratio
$R_{AA}$ become small ($\sim 10\%$ already at $p_{\perp}=10$ GeV)
\begin{table}
\caption{The suppression $R_{Au,Au}(p_{\perp})$ of neutral pions
at $\sqrt{s} = 200$ GeV. The $u_{min}$ and $z_{max}$ are shown
only for gluon jet.\label{Tab.2}}
\begin{tabular}{ccccc}
\hline \hline $p_{\perp}$  & $u_{\min}^g$ & $z_{\max}^g$ &
$R_{AA}$ &
\\ \hline
4 & 3 & 0.572 & 0.22
\\
 6 & 3.5 & 0.632 & 0.27
 \\
 8 & 4  &  0.667 & 0.34
 \\
10 & 4.3 & 0.7  &  0.38
\\
14 & 4.8 & 0.745 &  0.42
\\
20 & 5.4 & 0.786 &  0.44
\\
30 & 6  & 0.833  & 0.42
\\
40 & 6.2 & 0.866 & 0.38
\\
50 & 6.5 & 0.885 & 0.32
\\
60 & 6.8 & 0.898 & 0.25
\\
70 & 7.2 & 0.907 & 0.22

\\   \hline \hline
\end{tabular}
\end{table}

\begin{figure} [ht]
\centering \mbox{\includegraphics*[scale=1]{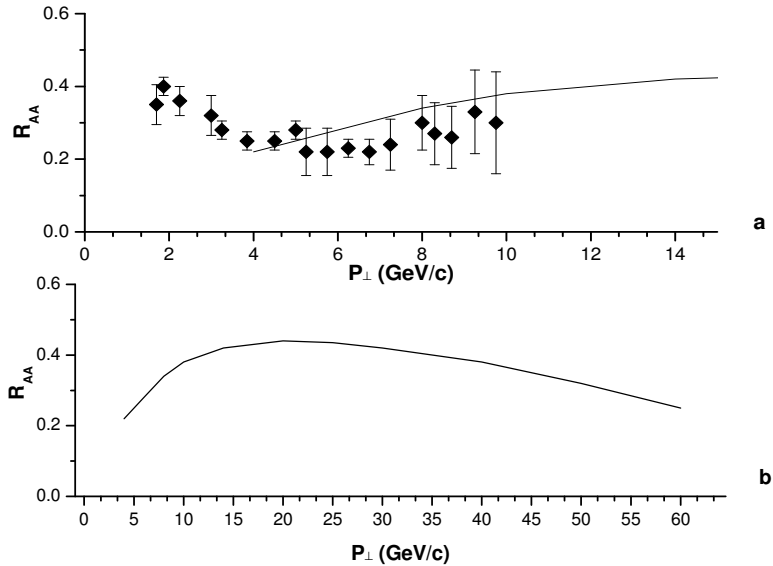}}
\caption{ a: The suppression factor $R_{AA}(p_{\perp})$ for
$\pi^{0}$ in central Au + Au at $\sqrt{s}= 200$ GeV in the range
of $p_{\perp}$ to 15 GeV/c and PHENIX data. b: The suppression
factor $R_{AA}$ which was calculated in the range of $p_{\perp}$
to 60 GeV/c. \label{Fig.3}}
\end{figure}

The dependence $R_{AA}(p_{\perp})$ is shown in Figure~\ref{Fig.3}.
This function have the weak rise above $p_{\perp} = 4$ Gev/c and
broad maximum in the range $p_{\perp} \sim 15-30$ Gev/c. The
PHENIX data there are only up 10 GeV/c~\cite{1}. It is possible to
describe these data by effective quasiparticle model with decrease
of effective coupling at $T \to T_c$ from above (look at
Table~\ref{Tab.1}) without additional free parameters. The
following decrease of $R_{AA}$ at $p_{\perp} > 20$ GeV is caused
by some factors --- in particular, by increase of $u_{\min}$ with
increase of $p_{\perp}$.

 It is interesting to note, that dependence $R_{AA}(p_{\perp})$ in
Figure~\ref{Fig.3} disagrees with monotonous increase at
sufficiently large of $p_{\perp}$, which was considered for
example in the paper~\cite{21}. This question have interest and
for $LHC$, where also  the continuous rise of suppression $R_{AA}$
with $p_{\perp}$ was found~\cite{19}. We will investigate this
question in the fortcoming work. But of course the calculation for
large $p_{\perp}$ may be questionable in very near region to
boundary of phase space $(p_{\perp}\simeq \sqrt{s}/2)$, where
overall energy and momentum conservation may be important.

\section{The calculations with accounting of intrinsic
transverse momenta of partons \label{sec4}}

The invariant cross section of hadron production in central $A+A$
collisions with accounting of intrinsic transverse momenta is
given by formula (\ref{eq.5}). We already have mentioned, that we
will consider for simplicity the collisions of flat disks. In this
case the integral of overlapping for $Au + Au$ collisions is $\bar
T_{AA}(b) = \frac{A^2}{\pi R_{A}^2} \simeq 262$ fm$^{-2}$. We
describe the Mandelstam variables for elementary parton-parton
scattering in term of light-cone variables $x_1$, $x_2$~\cite{8}:
\begin{eqnarray}
\label{eq.19}
 \hat s&=& x_{1}x_{2}\sqrt{s} +
 \frac{k_{\perp 1}^2 k_{\perp 2}^2}{x_{1}x_{2}\sqrt{s}}-2k_{\perp1} k_{\perp2}
\cos(\gamma_1 -\gamma_2), \nonumber \\
\hat t &=& -\frac{p_{\perp}}{z_c} \left(x_{1}\sqrt{s}\; e^{-y} +
\frac{k_{\perp 1}^2}{x_{1}\sqrt{s}} -2k_{\perp1}
\cos(\gamma_1)\right),
\nonumber\\
\hat u &=&-\frac{p_{\perp}}{z_c} \left(x_{2}\sqrt{s}\; e^{-y} +
\frac{k_{\perp 2}^2}{x_{2}\sqrt{s}} -2k_{\perp 2}
\cos(\gamma_2)\right). \
\end{eqnarray}
Here $p_{\perp}$ and y are the transverse momentum and rapidity of
produced particles, \mbox{$\cos(\gamma_1) = \frac{\bf{k}_{\perp
1}\bf{p}_{\perp}}
 {k_{\perp 1} p_{\perp}}$},
\mbox{$\cos(\gamma_2) = \frac{\bf{k}_{\perp 2}\bf{p}_{\perp}}
 {k_{\perp 2} p_{\perp}}$}. It is
usually assumed that initial $k_{\perp}$ distribution
$g_{N}(k_{\perp},Q^2)$ has a Gaussian form:
\begin{equation}
\label{eq.20} g_{N}(k_{\perp},Q^2) =
\frac{e^{\frac{-k_{\perp}^2}{<k_{\perp}^2>_{N}}}}{
\pi<k_{\perp}^2>_{N}}.
\end{equation}
In equation~(\ref{eq.5}) one should restrict the initial
transverse momentum $k_{\perp1} < x_{1}\sqrt{s}$ and
\mbox{$k_{\perp2} <x_{2}\sqrt{s}$}  such that longitudinal momenta
of partons have the same signs as their parent hadrons. We
introduce the new variables $t_1$ and$t_2$: $k_{\perp1} =
t_{1}x_{1}\sqrt{s}$, $k_{\perp2} = t_{2}x_{2}\sqrt{s}$, where $0
\le t_{i} \le t_{max}$ and $t_{max} < 1$. The variables
(\ref{eq.19}) one can describe in the following form:
\begin{eqnarray}
\label{eq.21} \hat s& =& sx_{1}x_{2} \Psi(t_{1},t_{2}, \gamma_{1},
\gamma_{2}), \nonumber\\
\hat t&=& -\frac{s x_{\perp}^{\pi}}{2z_c} x_{1}
\phi_{1}(t_{1},\gamma_{1}), \nonumber \\
\hat u &=& -\frac{s x_{\perp}^{\pi}}{2z_c} x_{2}\phi_{2}(t_{2},
\gamma_{2}),
\end{eqnarray}
where $x_{\perp}^{\pi} = \frac{2p_{\perp}^{\pi}}{\sqrt{s}}$.

Here we use designation:
\begin{eqnarray}
\label{eq.22}
\Psi(t_{1},t_{2}, \gamma_{1}, \gamma_{2}) &=& 1 +
t_{1}^{2}t_{2}^{2}-2t_{1}t_{2}cos(\gamma_{1}-\gamma_{2}), \nonumber\\
\phi_{1}(t_{1},\gamma_1) &=& 1 + t_{1}^2 - 2t_{1}cos(\gamma_{1}),
\nonumber \\
\phi_{2}(t_{2},\gamma_2) &=& 1 + t_{2}^2 - 2t_{2}cos(\gamma_{2}).
\end{eqnarray}
Now  for example instead of $f_{gg}(\xi)$ (\ref{eq.7}) we have:
\begin{eqnarray}
\label{eq.23}
f_{gg}(\xi,t_1,t_2,\gamma_{1},\gamma_{2}) =
\frac{3(\Psi\xi-\phi_2)}{\Psi^{2}\phi_{1}^{2}\xi^4}
-\frac{\phi_{1}\phi_2(\Psi\xi-\phi_2)^2}{\Psi^4 \xi^6\phi_{1}^3} +
\frac{\phi_2}{\xi^3(\Psi\xi-\phi_2)\Psi\phi_{1}^2} +
\frac{(\Psi\xi-\phi_2)^2}{\Psi \phi_2^{2}\phi_1^{2}\xi^3}.
\end{eqnarray}
We have analogous modification and for other elementary cross
section~(\ref{eq.8}-\ref{eq.10}).

In the structure function $G_{g}(x_1)G_{g}(x_2)$ we have the
modification: \mbox{$x_{1} = \frac{x_{\perp}^{\pi}\xi}{2z_c}$} and
\mbox{$x_{2} =
\frac{x_{\perp}^{\pi}\xi\phi_1}{2z_{c}(\Psi\xi-\phi_2)}$}.

The invariant cross section for hadron production (for example in
gluon jet) in central $Au + Au$ collisions with accounting of
partons initial momenta $k_{\perp}$ have the form:
\begin{eqnarray}
\label{eq.24}
E \frac{d\sigma^{AA}}{d^{3}p} &=& \frac{9 A^2}{\pi R_{A}^2}
\int\limits_{x_{\perp}^{\pi}}^{z_{max}^{g}(p_{\perp})}dz
\frac{z^{2}\alpha_{s}^{2}(Q(z))D_{h/g}(z_{c}^{\ast},{Q^{\ast}}^
2)}{p_{\perp}^4 (1 - \frac{\Delta E_{g}}{E}(z))}
\int\limits_{0}^{t_{max}(z)} dt_{1} t_{1}
\int\limits_{0}^{t_{max}(z)} dt_{2} t_{2} \nonumber\\
&& \int\limits_{0}^{2\pi}d\gamma_{1} \int\limits_{0}^{2\pi}d\gamma
_{2} \int\limits_{\frac{2\phi_{2}(t_2,\gamma_{2})z}
{2z\Psi(t_1,t_2,\gamma_{1},\gamma_{2})-\phi_{1}(t_1,\gamma_{1})x_{\perp}^{\pi}}}
^{\frac{2z}{x_{\perp}^{\pi}}}d\xi\,
g_{1}(t_1,\xi,z)g_{2}(t_1,t_2,\gamma_{1},\gamma_{2},\xi,z)\times \nonumber \\
&\times& f_{gg}(\xi,t_1,t_2,\gamma_{1},\gamma_{2})  x_{1}^{2}
x_{2}^{2} s^{2}
\Psi(t_1,t_2,\gamma_{1}\gamma_{2})x_{1}G_{g}(x_{1},Q^2)x_{2}G_{g}(x_{2},Q^2).
\end{eqnarray}
The value $z_{max}^{g}(p_{\perp})$ can be found from condition
$z_{max}^{g} = \frac{p_{\perp}}{p_{\perp}+
u_{min}^{g}(p_{\perp})}$, where the values
$u_{min}^{g}(p_{\perp})$ (and also $z_{max}^{g}$  which can be
found independently)
 there are in Table~\ref{Tab.2}.

In this work we neglect of the transverse momentum smearing from
the jet fragmentation~\cite{8}.

The Gaussians $g_1$ and $g_2$ in~(\ref{eq.24}) have the form:
\begin{eqnarray}
\label{eq.25}
&& g_{1}(t_1,\xi,z) = \frac{
\exp{\left(-\frac{t_{1}^{2}\left(\frac{x_{\perp}^{\pi}}{2z}\right)^{2}\xi^{2}s}
{<k_{\perp}^2>}\right)}} {\pi <k_{\perp}^2>}, \nonumber
\\
&& g_{2}(t_1,t_2,\gamma_{1},\gamma_{2},\xi,z) =  \frac{
\exp{\left(-\frac{t_{1}^{2}\left(\frac{x_{\perp}^{\pi}}{2z}\right)^{2}\xi^{2}s
\frac{ \phi_1^2(t_1,\gamma_1)}{
(\xi\Psi(t_1,t_2,\gamma_1,\gamma_2)-\varphi_2(t_2,\gamma_2))^2} }
{<k_{\perp}^2>}\right)}} {\pi <k_{\perp}^2>}.
\end{eqnarray}

With  accounting of nuclear shadowing the parton distribution
$G_g(x,Q^2)$ must be modified: $G_g(x,Q^2)S_{A}(x,Q^2)$.

In order to find the restriction $t_{max}(z)$, one should take
into account, that lower limit of integral on $\xi$ must be $>0$
at any values of t and $\gamma$. That correspond to condition:
$2\Psi_{min}z > \phi_{1 max} x_{\perp}^{\pi}$. But the value
$\Psi$ have minimum at $\gamma_{1} = \gamma_{2}$, and value
$\phi_{1}$ have maximum at $\gamma_{1} = \pi$, therefore we have
the condition : $2z(1+t)^{2}(1-t)^{2} > x_{\perp}^{\pi}(1+t)^{2}$,
i.e. $t < 1-\sqrt{\frac{x_{\perp}^{\pi}}{2z}}$. At mentioned
values of angles $\gamma$ the lower limit of $\xi$ is:
$\xi(z,t)=\frac{2z}{2z(1-t)^{2}-x_{\perp}^{\pi}}$, and we have $t
\le t_{max}$, where $t_{max}(z)$ correspond to $\xi(z,t_{max}) =
\xi_{max} = \frac{2z}{x_{\perp}^{\pi}}$. For set of values of
$p_{\perp}$ we calculate by numeral way the function $t_{max}(z)$.

For example, for $p_{\perp}$ = 8 GeV/c we find:
\begin{eqnarray} \label{eq.26}
&&t_{\max}(z) = 1 - \sqrt{\frac{0.04}{z}}-\nonumber\\ &-& 0.2929
\exp{(\sqrt{z-0.08})(0.0618 -10.9786z+27.8446z^2 - 29.0236z^3
+10.78z^4)}
\end{eqnarray}
 and for $p_{\perp}=20$ GeV/c we find correspondingly
\begin{eqnarray} \label{eq.27}
&&t_{\max}(z) = 1 - \sqrt{\frac{0.1}{z}} - \nonumber\\ &-& 0.2929
\exp{(\sqrt{z-0.2})(0.6808 -8.0142z + 15.8356z^2 - 14.1458z^3 +
4.744z^4)}.
\end{eqnarray}
At calculation of the invariant cross section by formula
(\ref{eq.24}) we use Monte-Carlo method. We use here the
factorization scale $Q = \frac{p_{\perp}}{2z_{c}}$ and
fragmentation scale $Q^{\ast} = \frac{p_{\perp}}{2z_{c}^{\ast}}$.
 In absence of
nuclear effects --- jet quenching and shadowing the equation
(\ref{eq.24}) describes the collection of binary NN collisions. We
have in this case $\Delta E = 0$ and $z_{max}^{q} = 1$. For binary
collisions we use the value $<k_{\perp}^{2}>_{pp}\simeq 1.8$
GeV$^2$~\cite{18}-\cite{19}.

In the presence of medium we use the estimation in Ref.~\cite{18}:
\begin{equation}
\label{eq.28}
<k_{\perp}^{2}>_{AA} = <k_{\perp}^{2}>_{pp} + C(\nu_{m} -1),
\end{equation}
where $C \simeq 0.4$, $\nu_{m} \simeq 4$, i.e. we have the
estimation $<k_{\perp}^{2}>_{AA} \simeq 3$. At calculation we use
also the estimation $<k_{\perp}^{2}>_{AA} \simeq 4$. This have a
weak influence on the value of suppression $R_{AA}(p_{\perp})$.

\begin{figure}[ht]
\centering \mbox{\includegraphics*[scale=1]{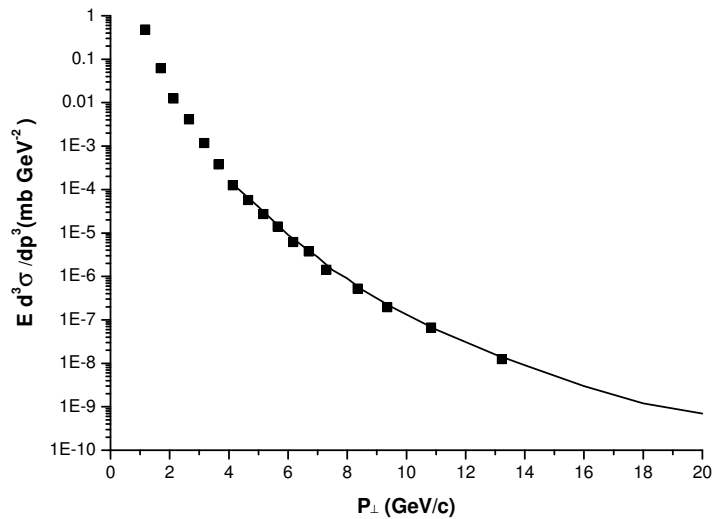}}
\caption{ Neutral pion spectra in $pp$ collisions  at $\sqrt{s}=200$
GeV. The data from PHENIX and the solid line is calculated
result. \label{Fig.4}}
\end{figure}

We use now the formula (\ref{eq.24}) for calculation of the
spectrum of high  $p_{\perp}$ neutral pions in $pp$ collisions at
$\sqrt{s}=200$ GeV. We take into account the sum of elementary
parton - parton cross section (\ref{eq.7}-\ref{eq.10}) with
accounting their modification of type (\ref{eq.23}). We have not
the divergence at calculation and therefore we do not introduce a
regulator $\mu^2$  in the denominators of elementary cross
section. We do not introduce also the K-factor. In
Figure~\ref{Fig.4} we show our  calculation for the spectrum of
high $p_{\perp}$ neutral pions in $pp$ collisions and PHENIX
data~\cite{20} up to $p_{\perp}\simeq 14$ GeV/c One can see  that
calculation by formula of type (\ref{eq.24}) for $pp$ collisions
reproduces the data in the region $p_{\perp} > 4$ GeV/c.

It should be noted, that we calculated also the factor
$R_{AA}^{\pi^0}(p_{\perp})$ at $\sqrt{s} = 200$ GeV by
formula~(\ref{eq.24}) with Cronin effect and shadowing, but
without the energy loss (for flat discs) using the value
$<k_{\perp}^2>_{AA}$ in the presence of medium (\ref{eq.27}). We
find the values $R_{AA} = 1.56$ at $p_{\perp} =4$ GeV/c and
$R_{AA}$ =1.1 at $p_{\perp} = 8$ GeV/c. That agrees with results
of paper~\cite{19}. We use also the formula (\ref{eq.24}) for
calculation of suppression $R_{AA}$ for great $p_{\perp}$ with
accounting of intrinsic momentum $k_{\perp,i}$. They give the
small increase of factor $R_{AA}$ - less than $10 \%$. The results
are shown in Table~\ref{Tab.2} and in Figure~\ref{Fig.3}.

It should be noted, that formula of type (\ref{eq.24}) can be used
also for investigation of the Cronin enhancement at SPS energy
($\sqrt{s}=17$ GeV).

\section{Conclusion \label{sec5}}

In this work we use the physical characteristics of initial and
mixed plasma phases which were found in the effective
quasiparticle model for investigation of suppression of neutral
pion spectra in central $Au + Au$ collisions at $\sqrt{s}=200$
GeV. The specific feature of this model consist in decrease of
thermal gluon mass and effective coupling in a region of phase
transition. We use  also the hypothesis of hot glue production at
the first stage of plasma expansion. We describe the following
evolution of plasma in isentropic model. The energy loss of the
high energy gluon and quark jets in plasma leads to suppression of
hadron spectra. For calculation of energy loss we use opacity
expansion~\cite{3}, which is dominated by first term. At
calculation we take into account the finite kinematic boundaries,
which often do not accounted for. We calculate the gluon density
in plasma and show that plasma is sufficiently thin. This to some
extent justify the model of the single hard medium-induced
scattering.

We show that suppression factor $R_{AA}(p_{\perp})$ has no
continuous rise with increase of $p_{\perp}$ - one has the weak
rise, then smooth maximum at $P_{\perp}\sim 20$ GeV/c and then it
decrease with further increase of $p_{\perp}$. This question is
interesting  for LHC, where the calculations~\cite{19,21} show
continuous rise. We will consider this question in the fortcomin
work. It is interesting to investigate also the hard photon
suppression - apparently it is much smaller than for
hadrons~\cite{17,21}.

We use in this work the simplified model of collisions of flat
discs at $\sqrt{s}=200$ GeV. We are planning to use the more
complex model of nuclear geometry in the forthcoming work. It
should be noted, that model of collisions with transverse
expansion and Woods-Saxon nuclear distribution with some
parametrization of energy loss was used in the work~\cite{9}.
There was found also slight increase of nuclear modification
factor $R_{AA}$ with increase of $p_{\perp}$ at $200$ GeV (at
  accounting of both gluon and quark jets). In recent work we used
the parameters  which were found in effective quasiparticle model
and nuclear modification of structure function from~\cite{12}.  We
take into account also the intrinsic transverse momentum of
partons $k_{\perp,i}$. We show, that one can avoid of divergence
at large $k_{\perp,i}$, if we take into account the physical
restrictions on initial transverse momentum. Therefore there is no
necessity to introduce a regulator $\mu^{2}$ in parton-parton
cross-section. It can have the meaning at more low SPS energies,
for example at investigation of Cronin effect, where the
suppression factor $R_{AA}$ can be much larger than 1.

\begin{acknowledgements}
The authors thank Profs.\ S.\ T.\ Belyaev, V.\ I.\ Manko for
fruitful discussion. The authors are grateful to Prof.B.\ V.\
Danilin and Dr.\ A.\ V.\ Lomonosov for a careful reading of the
manuscript. The work was supported by the grant PU-112/001/121 of
the Russian Ministry of Industry and Science.

\end{acknowledgements}

\end{document}